\documentclass{aa}  
\usepackage{textcomp}

\usepackage{graphicx}
\usepackage{txfonts}
\usepackage{natbib}
\usepackage{xcolor}
\usepackage{gensymb}

\begin{document}

   \title{EWOCS-VIII: Internal kinematics and expansion of Westerlund 1 from VVVX proper motions\thanks{Corresponding author: \email{ccordenes@uc.cl}}}
   \titlerunning{EWOCS-VIII}

   \author{C. Ordenes-Huanca
          \inst{1,2}, A. Bayo\inst{3}, M. Guarcello\inst{4}, M. Zoccali\inst{1},
          A. Rojas-Arriagada\inst{5,6}, Nicholas J. Wright\inst{7},
          S. Sciortino\inst{4},
          F. Damiani\inst{4},
          K. Muzic\inst{8},
          V. Almendros-Abad\inst{4},
          M. Andersen\inst{3},
          R. Bonito\inst{4},
          M. Haberle\inst{3},
          T. J. Haworth\inst{9}
          \and 
          L. Prisinzano\inst{4}
          }
    \authorrunning{Ordenes-Huanca et al.}

   \institute{Instituto de Astrofísica, Pontificia Universidad Católica de Chile, Casilla 306, Santiago 7820436, Chile
        \and
            Departamento de Astronomía, Universidad de Concepción, Casilla 160-C, Concepción, Chile
        \and
             European Southern Observatory, Karl-Schwarzschild-Strasse 2, D-85748 Garching bei München, Germany
        \and
            Istituto Nazionale di Astrofisica (INAF) – Osservatorio Astronomico di Palermo, Piazza del Parlamento 1, 90134 Palermo, Italy
        \and
            Departamento de Física, Universidad de Santiago de Chile, Av. Víctor Jara 3659, 9170124, Santiago, Chile
        \and
            Center for Interdisciplinary Research in Astrophysics and Space Exploration (CIRAS), Universidad de Santiago de Chile, Santiago, Chile
        \and 
            Astrophysics Group, Keele University, Keele ST5 5BG, United Kingdom
        \and
            Instituto de Astrofísica e Ciências do Espaço, Faculdade de Ciências, Universidade de Lisboa, Ed. C8, Campo Grande, 1749-016 Lisbon, Portugal
        \and
            Astronomy Unit, School of Physics and Astronomy, Queen Mary University of London, London E1 4NS, United Kingdom
             }

   \date{Received June 4, 2026; accepted July 29, 2026}

 
  \abstract
   {Westerlund 1 (Wd1) is the most massive young star cluster known in the Milky Way and a key laboratory for studying the early dynamical evolution of massive clusters. Owing to its high extinction, the internal kinematics of its intermediate-mass stellar population remain largely unexplored.}
   {We aim to characterize the internal kinematic properties of Wd1 using a homogeneous census of near-infrared (NIR) selected cluster members and their proper motions (PMs). We considered PMs from the VIRAC2 catalog, based on multi-epoch VVV/VVVX observations.}
   {For $1286$ candidate members previously identified through NIR photometry and astrometry (Paper I), we computed their PMs relative to the cluster mean motion and analyzed them as a function of position and radius. We also investigate radial trends, test the robustness of the results against the assumed cluster center, and search for preferred directions of motion.}
   {We detect a statistically significant signature of expansion in the outer regions of Wd1, with the radial component of the relative PMs increasing with distance from the cluster center. The expansion appears asymmetric, with the strongest gradient detected along a PA$=84\degree \pm 8\degree$ in the plane of the sky and with $5.9\sigma$ significance. We also find hints for inward radial motions in the central region, consistent with ongoing mass segregation, but only at $\approx 2\sigma$ significance. Taken together, these results are consistent with a nearly monolithic formation scenario.}
   {}


   \maketitle
%

\section{Introduction}

Stellar kinematics encode a fossil record of the conditions under which a cluster formed. Whether the latter collapsed violently from a single gas concentration or grew through the merger of smaller subclusters should still be legible in how its members move today \citep{Allison_2009, Arnold_2022}. Recovering these signatures requires astrometry that extends beyond the bright massive stars that Gaia has observed \citep{gaiadr3_pm} and spatial coverage that extends beyond the dense cluster cores, as observed by the limited field-of-view ($\approx 3.5' \times 3.5'$) of the Hubble Space Telescope (HST) and its Advanced Camera for Surveys (ACS) Wide Field Camera
(WFC; \citeauthor{Gonzaga_2011} \citeyear{Gonzaga_2011}).\\

The formation of open clusters remains a subject of debate, with two main scenarios proposed in the literature. In the monolithic formation scheme, clusters form in situ within a single, dense gas concentration, resulting in centrally dense systems with smooth radial stellar distributions \citep{Banerjee_2015}. In very young massive clusters ($\gtrsim10^{4} M_{\odot}$, ages of $\lesssim 5$ Myr), stars are still embedded in their natal molecular cloud. Feedback from massive stars, through radiation and outflows, can rapidly expel the residual gas, leaving the stellar system out of virial equilibrium and potentially triggering a phase of violent expansion \citep{Lada_1984}.\\

Alternatively, the hierarchical formation scenario proposes that massive clusters assemble through the merger of smaller subclusters, reflecting the substructured nature of dense gas in molecular clouds \citep{Longmore_2014, Wright_2019}. In this scenario, the surrounding molecular gas enhances gravitational infall, accelerating the merging process \citep{Longmore_2014}. Although direct observations of cluster mergers remain scarce \citep{Kuhn_2019}, this mechanism is expected to operate primarily at early and deeply embedded stages of cluster formation.\\

Observationally, OB associations and young massive clusters often exhibit spatial substructure, as well as a configuration consisting of a dense core surrounded by a more extended halo \citep{Kuhn_2014}. Even if the merging phase itself is no longer observable, kinematic imprints of these early processes may persist. In particular, proper motion (PM) measurements can reveal signatures of past collapse or expansion linked to the history of the formation of the cluster. Such kinematic signatures may include not only global expansion, but also differential motions such as central contraction associated with mass segregation or subvirial states \citep{Cottaar_2012}.\\

The proper motions (PMs) provided by Gaia DR3 \citep{gaiadr3_pm} are a fundamental tool for identifying members of the Westerlund~1 (Wd1) cluster and for characterizing its mean bulk motion \citep{Clark_2010, Negueruela_2022}. Gaia data releases \citep{Gaia_2018, Gaia_2023} have delivered photometry and astrometry that constrain some of the most fundamental parameters of the cluster, including its distance and age. For instance, using Gaia DR2 data and a Bayesian approach, \citet{Aghakhanloo_2020} placed Wd1 at $2.6$~kpc with $18\%$ precision, though they cautioned that this estimate may be affected by the Gaia parallax zero-point offset. In contrast, \citet{Negueruela_2022} derived a substantially larger distance of $4.23$~kpc from Gaia EDR3 parallaxes, in better agreement with earlier studies, like the $\approx3.8$~kpc distance found by \citet{Lim_2013} and by \citet{Davies_2019} using NIR and Gaia DR2 data, respectively.\\

Besides its distance, several studies seem to converge towards an age for Wd1 of about $\approx 5$--$6$ Myr. Optical and NIR data have allowed us to obtain such values \citep{Clark_2005, Negueruela_2010, Kudryavtseva_2012, Lim_2013}. Nevertheless, more recent studies have suggested an older age for part of the Wd1 population. \citet{Beasor_2021} obtained a value of $10.4$ Myr, using mid-IR data of cool supergiant members of the cluster, and $7.2$ Myr for its pre-Main Sequence stars. \citet{Navarete_2022}, on the other hand, computed an age of $10.7\pm1$ Myr for its red supergiants, using Gaia EDR3 combined with NIR and far-IR data. Both studies therefore suggest that Wd1 experienced several episodes of star formation. Such discrepancies are unsurprising given that Wd1 suffers from extreme foreground extinction ($A_{V} \approx 10$~mag), which severely limits optical detections and introduces uncertainties in the derived parameters.\\

Beyond Gaia data, the work by \citet{Wei_2025} analyzed HST WFC3-near infrared (NIR) astrometric and photometric data for stars within a $\approx 3.5' \times 3.5'$ region centered on Wd1. They derived membership probabilities based on both PMs and photometric colors, and identified $3\,586$ stars with membership probabilities $\geq 0.3$ and in the mass range between $\approx 1.1$--$15.2$~$M_{\odot}$. These sources showed that the cluster is subvirialized, with weak signs of mass segregation. Their study provides crucial insights into Wd1's central population, but its spatial coverage is restricted to the inner $\approx 12$ arcmin$^2$, leaving the kinematic properties of the extended cluster region essentially unconstrained.\\

In \citeauthor{Ordenes-Huanca_2026} (\citeyear{Ordenes-Huanca_2026}, from now on Paper I), we exploited the potential of the VISTA Variables in the Vía Láctea survey and its extension (VVV and VVVX, \citeauthor{minniti+2010} \citeyear{minniti+2010}) to unveil the intermediate-mass population ($1.5$--$20$ $M_{\odot}$) of Wd1. Using features extracted mostly from the VIRAC2 catalog \citep{Smith_2025}, we employed HDBSCAN to separate the Wd1 population from foreground and background contaminants. This work follows the more detailed analysis of kinematic trends and/or substructures in the candidate members, and the constraints these can place on Wd1’s history. VIRAC2 provides multi-epoch point-spread-function (PSF) photometry and PMs for more than $545$ millions sources in the Galactic plane, enabling the measurement of accurate PMs for faint and highly reddened stars that are inaccessible to optical surveys. Within the framework of the Extended Westerlund Open Clusters Survey (EWOCS\footnote{\url{https://westerlund1survey.wordpress.com/}}), these data offer a unique opportunity to probe the internal kinematics of intermediate-mass members of Wd1, complementing Gaia measurements for the brightest stars \citep{Negueruela_2022} and JWST/X-ray studies at lower masses \citep{Guarcello_2024, Guarcello_2025}.\\

This paper is structured as follows. Sec.~\ref{validation} includes the comparison between the VIRAC2 PMs of our member selection with those obtained with optical data from Gaia and HST. The kinematic features obtained through the PMs of our census are analyzed in Sec.~\ref{kinematics}. We discuss the implications of an expansion in Wd1 in Sec.~\ref{discussion} and summarize our conclusions in Sec.~\ref{conclusions}.\\


\section{Validation of VIRAC2 PMs}
\label{validation}

In paper I, we identified $1286$ intermediate-mass members of Wd1 using a {\tt HDBSCAN} clustering algorithm \citep{Campello2013, McInnes2017}. The latter considered six input parameters obtained from the VIRAC2 source catalog \citep{Smith_2025}: the equatorial coordinates $\alpha$ and $\delta$, equatorial PMs $\mu_{\alpha *}$ and $\mu_{\delta}$, $J-K_{\rm s}$ color, and the minimum distance between each source and the $5$ or $6$ Myr PARSEC isochrone on the $J$ vs. $J-K_{\rm s}$ color-magnitude diagram.\\

To further validate the membership identified in Paper I, we cross-matched the $1286$ proposed members catalog from Paper I with the Gaia Data Release 3 main source catalog \citep{GaiaCollaboration_2016, Gaia_2023}. For the matching, we used a $1''$ radius and kept only the closest counterpart within it. The result is a catalog of $816$ common sources. The mean PM values of our $1286$ member candidates are $\overline{\mu_{\alpha*}}=-2.36 \pm 0.02$ mas$/$yr and $\overline{\mu_{\delta }}=-3.70 \pm 0.02$ mas$/$yr, where the uncertainties are simple standard errors of the mean. From our census, the Gaia data contain kinematic information for only $701$ stars.\\

\begin{figure*}
    \centering
    \includegraphics[width=0.8\textwidth]{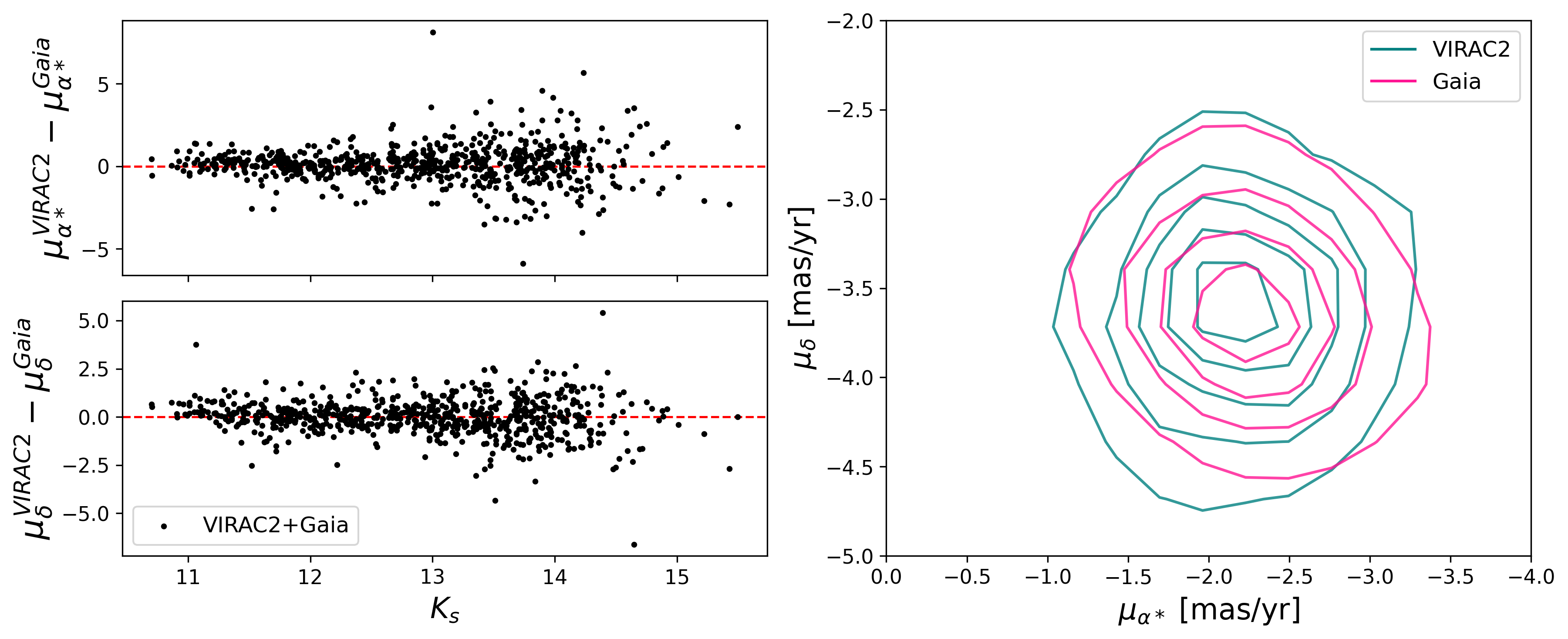}
   \caption{\textit{Left panels:} Equatorial PM values comparison between VIRAC2 and Gaia cross identified stars vs. their $K_\mathrm{s}$ magnitude. RA (Dec) difference is shown in the upper (lower) left panel and the dashed red line indicates zero difference. \textit{Right panel:} VPD for the common stars between VIRAC2 and Gaia DR3 datasets. Teal contours are VIRAC2 values, whereas magenta ones are Gaia measurements.}
    \label{virac2_gaia}%
\end{figure*}

The difference between VIRAC2 and Gaia PM measurements for each equatorial coordinate is shown in the upper and lower left panels of Fig.~\ref{virac2_gaia}, where the horizontal axis represents the mean $K_\mathrm{s}$ magnitude for each star in common. One can observe good agreement between both datasets, at least for brighter stars, in which the standard deviation of the difference of those between $11 \leq K_\mathrm{s} \leq 13$ is about $0.7$ mas/yr in both equatorial coordinates. The difference in PM values is more evident for the fainter ones, as expected. The standard deviation for stars in the range $13 \leq K_\mathrm{s} < 15$ is about $1.4$ mas/yr for the $\alpha*$ component and $1.2$ mas/yr for the difference in the $\delta$ component. The right panel of Fig.~\ref{virac2_gaia} shows the vector point diagram (VPD) of the stars in common, using VIRAC2 PM measurements (in teal) and Gaia values (in magenta). Again, the contours show excellent agreement between both datasets. The summary of all the comparisons mentioned in the text is in Table~\ref{tab_comparison}.\\

\begin{table}[h]
\caption{Summary of the mean equatorial PM measurements, $\overline{\mu_{\alpha*}}$ and $\overline{\mu_{\delta}}$, for the different datasets mentioned in the text.}
\centering
\begin{tabular}{l|c|c}
\hline\hline 
\multicolumn{1}{c|}{Dataset}       & $\overline{\mu_{\alpha*}}$ {[}mas$/$yr{]} & $\overline{\mu_{\delta}}$ {[}mas$/$yr{]} \\ \hline \hline
All VIRAC2 candidates              & -2.36 $\pm$ 0.02                                    & -3.70 $\pm$ 0.02                                    \\
Bona fide members$^{a}$ & -2.231 $\pm$ 0.008                                   & -3.697 $\pm$ 0.008                                   \\
VIRAC2+Gaia$^{b}$                        & -2.25 $\pm$ 0.02                                    & -3.62 $\pm$ 0.02                                    \\
Gaia DR3$^{c}$                           & -2.35 $\pm$ 0.04                                    & -3.63 $\pm$ 0.03                                   \\ \hline \hline

\end{tabular}
\tablefoot{
\tablefoottext{a}{Measurements from \citet{Negueruela_2022}, comprising $401$ stars.}
\tablefoottext{b}{VIRAC2 candidates in common with Gaia DR3 with NIR PMs.}
\tablefoottext{c}{VIRAC2 candidates in common with Gaia DR3 with optical PMs.}
The latter two rows comprise $701$ sources, each with its standard error of the mean.}
\label{tab_comparison}
\end{table}

Table~\ref{tab_comparison} also shows the PM values obtained by \citet{Negueruela_2022}, which considered Gaia EDR3 data to obtain $401$ bona fide members. Within the errors, there is good agreement between the values of our candidates and those of Gaia. \\

\begin{figure}
    \centering 
    \includegraphics[width=0.4\textwidth]{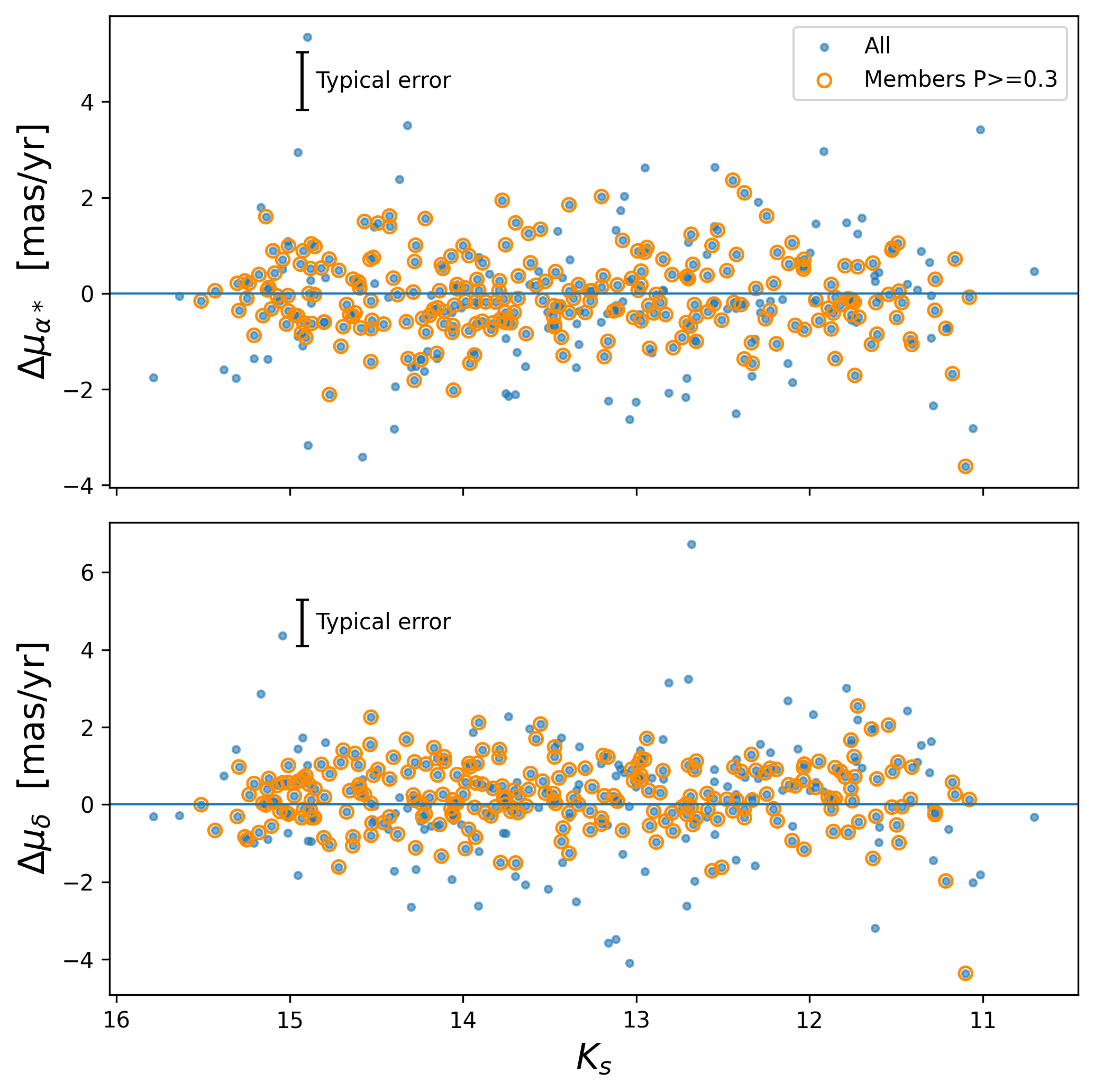}
   \caption{\textit{Top panel:} Difference between the RA component of raw VIRAC2 and \citet{Wei_2025} PMs, $\Delta \mu_{\alpha*}$, against $K_\mathrm{s}$ magnitude. Common sources are in blue, and those in orange are identified as members in both datasets. \textit{Bottom panel:} Same as the top panel, but for the Dec component of PMs, $\Delta \mu_{\delta}$. Typical errors in each PM coordinate are shown in both panels.}
    \label{OH_Wei_pms}%
\end{figure}

More recently, \citet{Wei_2025} presented a list of more than $10\,000$ stars in the line-of-sight of Wd1 and finding, among them, $3500$ Wd1 members. Their selection relied on HST optical and NIR data, using photometric colors and PMs. To compare the datasets, we cross-matched our data with their entire sample and found $374$ common sources. Considering their defined membership probability threshold of $P\geq 0.3$, we found that $250$ of these matches are labeled as members by their analysis. We computed, using the data of Table \ref{tab_comparison},
the difference between the mean equatorial PMs of our Wd1 member list and of the \citet{Wei_2025} member list. After correcting their PM values for the derived shift, we computed the difference between the VIRAC2 PMs and the
corrected \citet{Wei_2025} PMs. Those differences are shown as a function of $K_\mathrm{s}$ in Fig.~\ref{OH_Wei_pms} for RA (top panel) and Dec (bottom panel) coordinates.\\

To check if those differences are compatible with PM uncertainties, we have computed the typical error, shown in Fig.~\ref{OH_Wei_pms}, as the median of the sum in
quadrature of individual error, and its size accounts for most of the PM differences shown in Fig.~\ref{OH_Wei_pms}. The $76\%$ of the stars selected as Wd1 members by both studies (orange open circles) are within $2\sigma$ PM errors in both coordinates, thus, showing good agreement. The median PM differences are $\Delta \mu_{\alpha *}= -0.14$ mas/yr and $\Delta \mu_{\delta}= -0.20$ mas/yr. Sources with higher PM discrepancies are mostly identified as members only by our study. Considering stars with similar PM values that were not selected by the HST data, photometry is the main origin of their rejection, since \citet{Wei_2025} uses the distance between a given star to the assumed isochrone ($7.23$ Myr and $3.7$ kpc) in the $F160W\mathrm{mag}$ vs. $F125W\mathrm{mag}-F160W\mathrm{mag}$ CMD as a selection constraint. We attribute this discrepancy to the age and distance assumed in our work ($5$--$6$ Myr and $4.23$ kpc) compared to those reported by \citet{Wei_2025} .\\

\begin{figure*}[h]
    \centering 
    \includegraphics[width=0.8\textwidth]{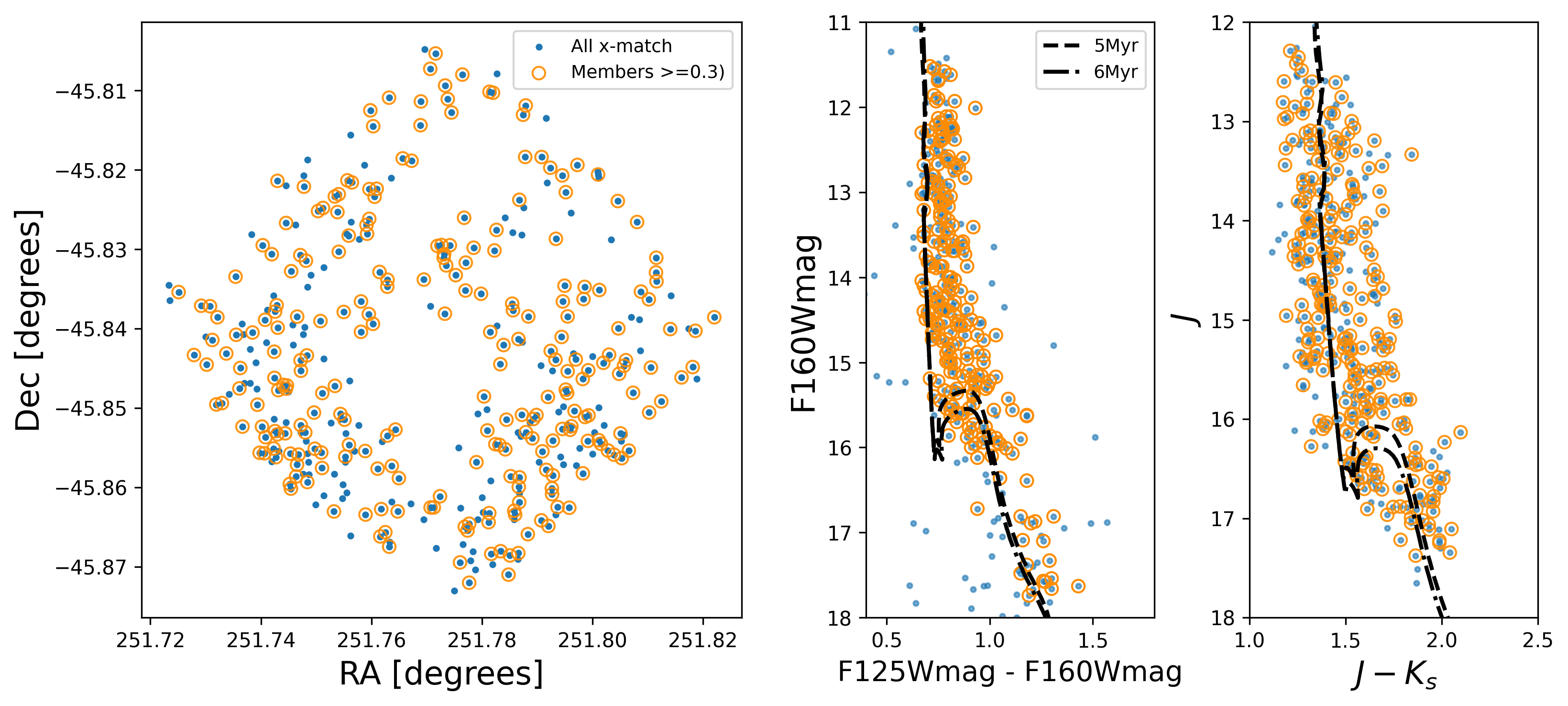}
   \caption{\textit{Left panel:} Spatial distribution of all common stars (blue dots) between this study and the one by \citet{Wei_2025}. The ones marked with an orange circle are identified as Wd1 members in both. The central inner region here is devoid of star counts due to the presence of saturated stars not included in the VIRAC2 source catalog. \textit{Central panel:} HST $F160W$ vs. $F125W-F160W$ CMD for the same subsets of stars. \textit{Right panel:} VIRAC2 $J$ vs. $J-K_\mathrm{s}$ CMD with stars color-coded as in the other two panels. $5$ (dashed line) and $6$ Myr (dash-dotted line) isochrones are also shown in both CMDs.}
    \label{OH_Wei}%
\end{figure*}

In addition to the age and distance assumed, another discrepancy can be due to their selection being made in the inner part of Wd1, comprising the central $\approx 3.5' \times 3.5'$, which is a much smaller area than our almost $10'$ circular region. Another source of discrepancy can arise from the absence of star counts in the very inner parts of the cluster, due to VVVX/VIRAC2 saturation from the massive stars in that region. The spatial distribution of all $374$ common sources between their and our selection (blue dots) is shown in the left panel of Fig.~\ref{OH_Wei} along with the ones identified as Wd1 members in both datasets (open orange circles). The central panel of Fig.~\ref{OH_Wei} shows a CMD in HST passbands, along with isochrones for the ages considered in our study, assuming a distance of $4.23$ kpc. The faint population in this diagram (between $17\leq$ F160Wmag $\leq 18$ and $0.5\leq$ F125Wmag$-$F160Wmag $\leq 1.0$) shows that our sample contains Wd1 members not identified by \citet{Wei_2025}. Their lower magnitudes result in both lower-quality (greater uncertainty) PM measurements and photometry, making their color bluer than predicted from the isochrone, likely accounting for their not being recognized as Wd1 members by \citet{Wei_2025}.\\

\section{Kinematic analysis}
\label{kinematics}

PMs have been used as a tool to identify expansion effects on open clusters (see e.g. \citeauthor{Wright_2024} \citeyear{Wright_2024}), these kinds of analyses are prone to projection effects. The radial motion of the cluster can induce an artificial expansion signature, so we corrected our PMs by radial velocity following \citet{Kuhn_2019} and using the mean RV value of $-20.36$ km s$^{-1}$ found by \citet{Tarricq_2021} for Wd1 members.\\

\begin{figure*}
    \sidecaption 
    \includegraphics[width=0.45\textwidth]{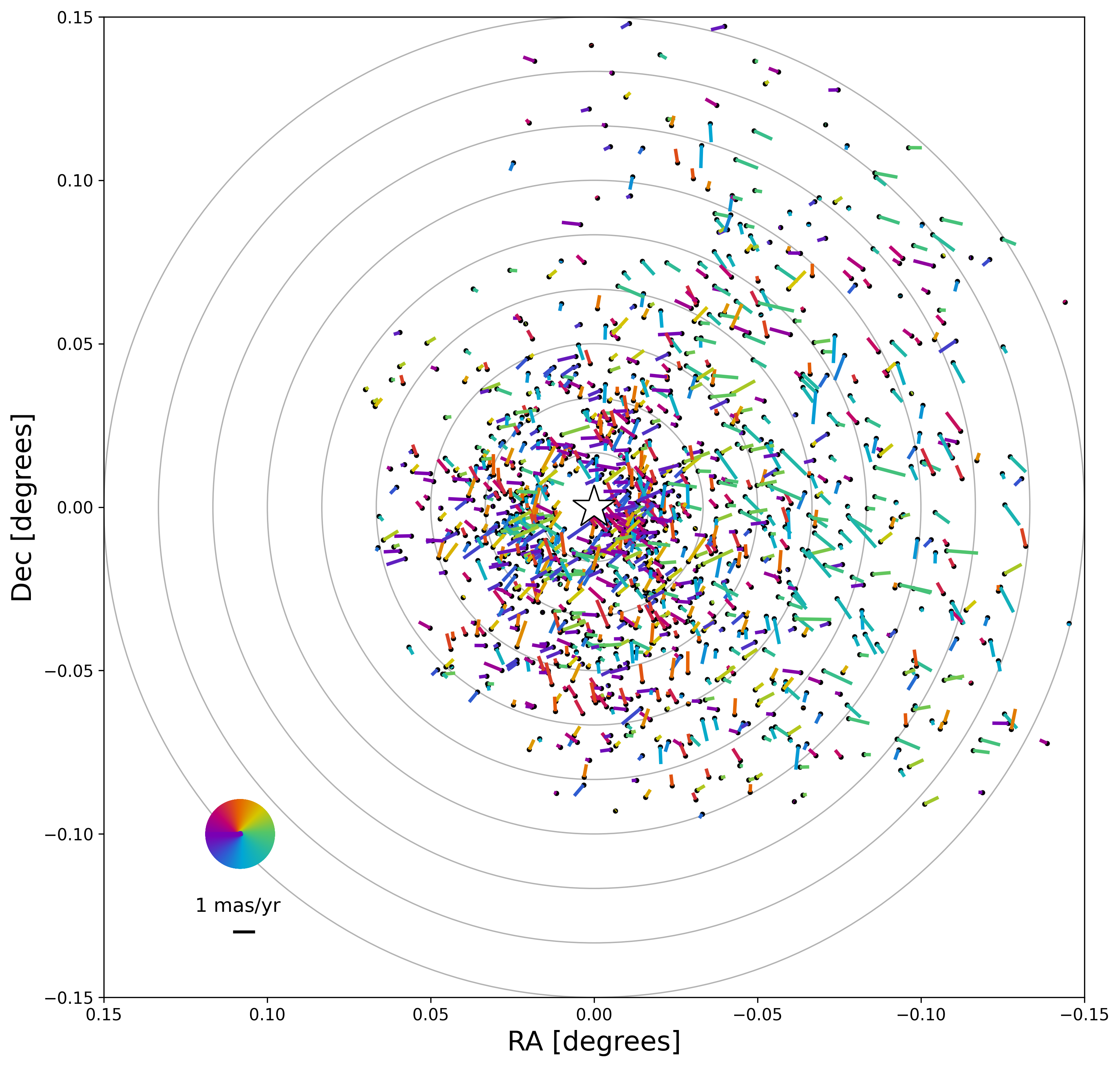}
   \caption{Spatial distribution of our candidate members (black dots) with their relative PM vectors (RV corrected) extending from them. The color of the PM vector indicates its direction on the color wheel in the lower left of the plot. Concentric circles have radii from $1'$--$9'$ in steps of $1'$. The scale of the PM vectors is indicated below the color wheel, and the black empty star symbol marks the center of the cluster.}
    \label{PM_color}%
\end{figure*}

Considering the radial velocity corrected PMs of all our member candidates and having the mean PM of the cluster, we derived the relative PM of each star: $\mu_{\alpha*,rel}=\mu_{\alpha*} - \overline{\mu_{\alpha*}}$ and $\mu_{\delta,rel}=\mu_{\delta} - \overline{\mu_{\delta}}$. Following \citet{Wiesneth_2025}, the relative motion of each object in the plane of the sky is shown as a line in Fig~\ref{PM_color}, where its direction is color coded and pointing as indicated by the color wheel in the lower left part of the plot. The considered PM scale is also shown. A concentration of light blue and light green lines are located towards the right side of the plot, suggesting that these stars are moving away from the cluster center. On the other hand, more purple lines are located on the left side of the same plot, suggesting they move in the opposite direction of those colored light blue and light green, as indicated by the color wheel. These hints of opposite directions of relative movements are not uncommon and show that expansion is a frequent behavior of most open clusters and OB associations \citep{Wright_2024} .\\

\begin{figure}
    \centering  \includegraphics[width=0.46\textwidth]{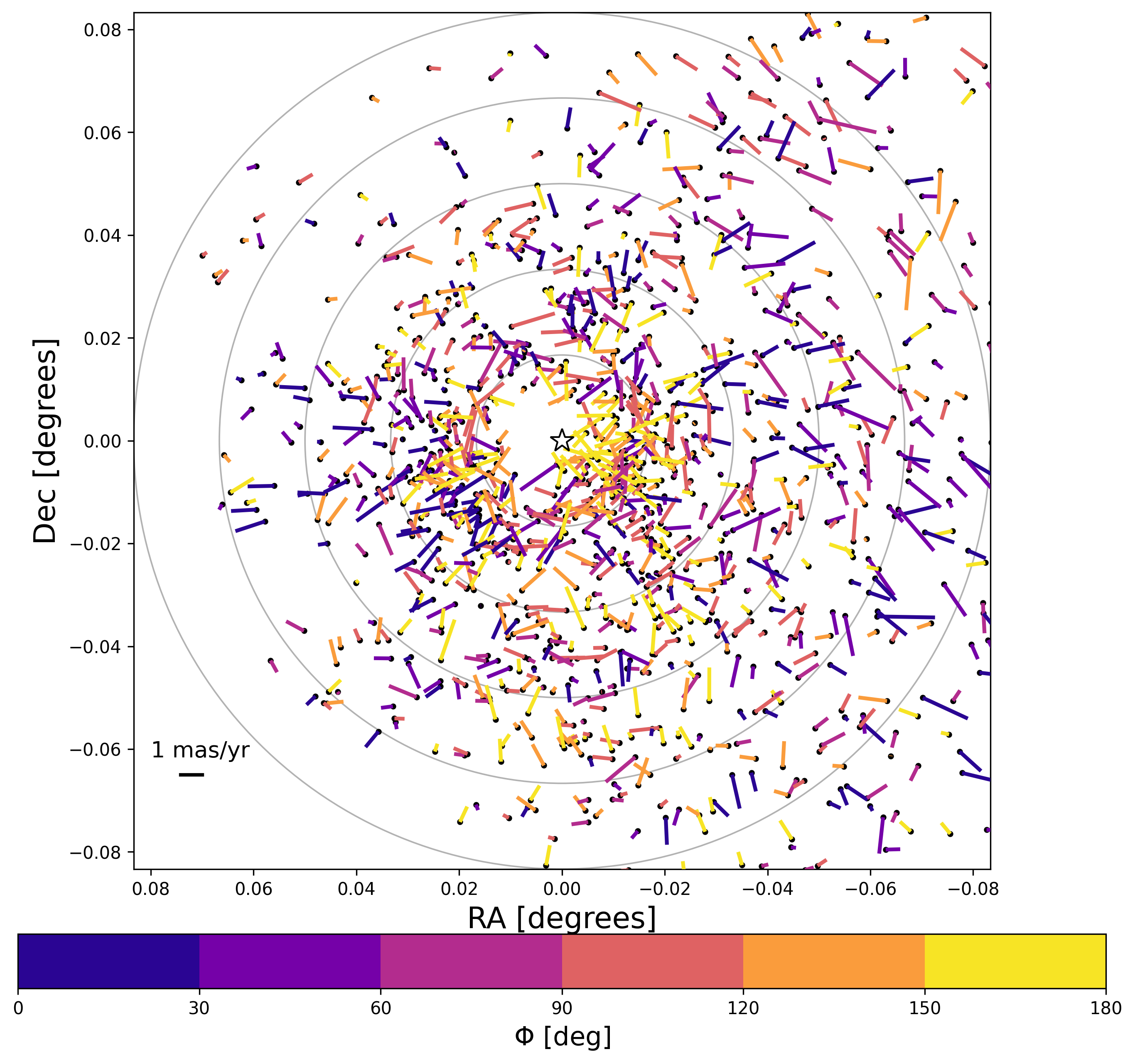}
   \caption{Zoom into the center of the spatial distribution of our candidate members (black dots). The PM vectors are RV corrected and colored according to the angle between $\mu_{r}$ and the line that connects each star with the center of the cluster (marked with a black empty star). Concentric circles have radii from $1'$--$5'$ in steps of $1'$. Purple colored PM vectors show stars moving away from the center, while PM vectors in yellow show stars moving towards the center.}
    \label{PM_phi}%
\end{figure}

In order to confirm if this behavior is related to signs of cluster expansion, we computed the angle between the PM vector of each star and the line that connects the center of Wd1 with its position, namely $\Phi$. The latter helps to visualize the kinematic behavior because stars moving away from the cluster center will show $\Phi \approx 0\degree$, whereas those approaching the center will have $\Phi \approx 180 \degree$. Tangential movements will have $\Phi \approx 90\degree$. Fig~\ref{PM_phi} shows a zoom into the inner $5' \times 5'$ region of Wd1, where we have plotted their relative PM values, as in Fig~\ref{PM_color}, but now each line is colored according to its $\Phi$ value. The plot shows that stars in the outer parts are moving mostly away from the Wd1 center (purple lines with $0\degree\leq \Phi \leq 30\degree$), suggesting a degree of expansion among the selected member candidates. On the other hand, a concentration of yellow lines near the center of the cluster suggests that part of the central stars are moving towards the center of Wd1. We note here that $\Phi$ was computed using a small-angle approximation. Nevertheless, we recomputed it using spherical sky geometry and found that the differences are negligible for the angular extent considered here, with a median absolute difference of $\sim0.018$ degrees. The latter is a much smaller value compared to the $\Phi$ bins considered in Fig.\ref{PM_phi} (of $30$ degrees), so the results remained unchanged.\\

Therefore, to further analyze this possible kinematic behavior, we also computed the radial component of the relative PM vectors of the stars, $\mu_{r}$. The definition of this parameter is the projection of the relative PM vector onto a line connecting the the cluster center to the position of each star (white star and black dots in Fig.~\ref{PM_color} and Fig.~\ref{PM_phi}, respectively). For positive (negative) $\mu_{r}$ values, the star would be moving away from (towards) the cluster center \citep{Wiesneth_2025}. \\

\begin{figure}
    \centering  \includegraphics[width=0.45\textwidth]{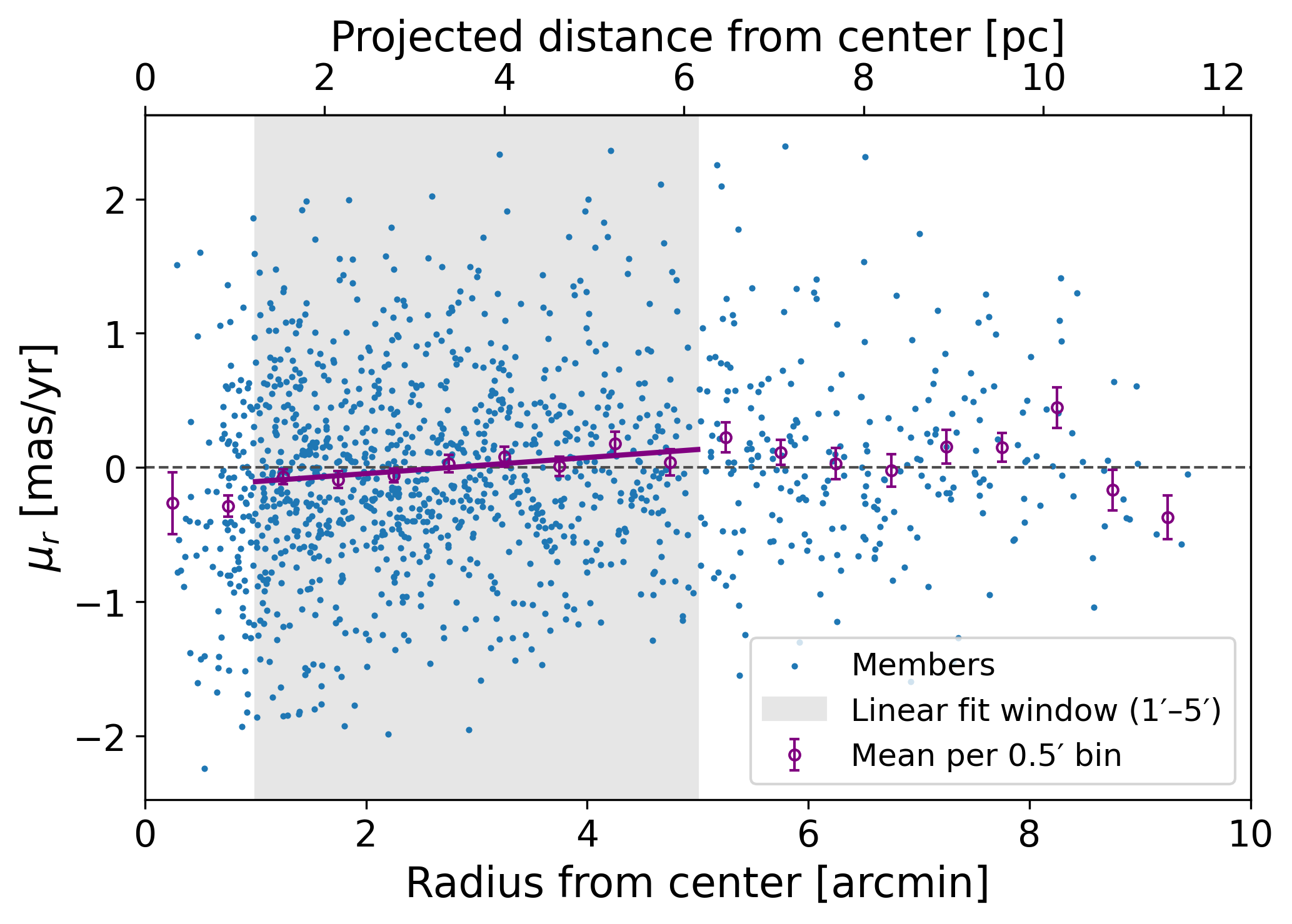}
   \caption{Relative radial PM vector, $\mu_{r}$, RV corrected, against the radial distance from the cluster center (bottom axis) and the projected radius in pc, according to the cluster distance (top axis), of each of our member candidates (blue dots). Purple dots show the mean $\mu_{r}$ value in bins of $0.5'$. Error bars correspond to the standard error of the mean. The gray-shaded region indicates the data used for the linear fit, depicted as a purple line.}
    \label{radial_PM_fit}%
\end{figure}

The relative radial PM, $\mu_{r}$, value is plotted against the distance of each star from the cluster center in Fig~\ref{radial_PM_fit}. Blue dots indicate these parameters for our $1286$ candidates, and those in purple mark the mean $\mu_{r}$ for $0.5'$-wide bins, with the bars corresponding to the standard error of the mean. Since the mean $\mu_{r}$ can be affected by the number of points in each bin, we focused on the region within $1'$--$5'$, which is more populated. Here, we made a linear fit, obtaining a slope of $0.06 \pm 0.02$ mas yr$^{-1}$ arcmin$^{-1}$ where a bootstrap resampling of stars within radial bins estimates the reported error. An additional systematic uncertainty associated with the adopted cluster center is estimated by repeating the fit for centers displaced by $30''$ in multiple directions, as described below. We found that the intercept of the fit is $-0.16 \pm 0.05$ mas yr$^{-1}$. This intercept indicates that the possible expansion of outermost candidates can also be observed from this parameter, at least up to $5'$, where we have more data points. In addition, a possible contraction of the central part of the cluster, at radii within $2'$ is observed. The linear fit also shows that the horizontal intercept is $2.5'\pm0.85'$; thus, the possible contraction extends up to this radial distance from the cluster center.\\

\subsection{Robustness of expansion vs. cluster center reference}

\begin{figure*}
    \centering  \includegraphics[width=0.75\textwidth]{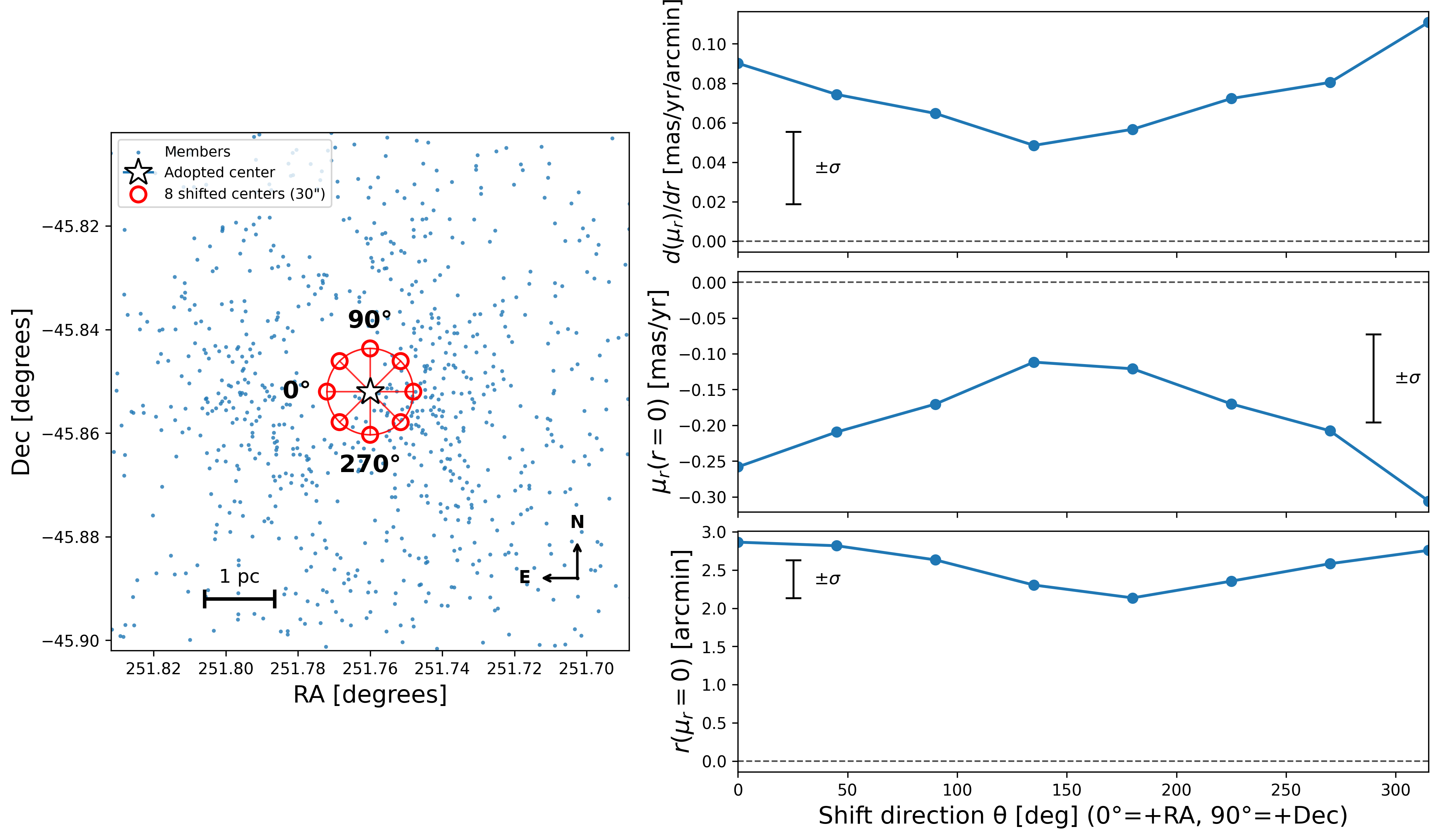}
   \caption{\textit{Left panel:} Zoom into the central region of the spatial distribution of our Wd1 members selected in Paper I. Different locations of the center for computing the $\mu_{r}$ values of our proposed Wd1 members are marked with red empty circles. The empty star marks the original adopted center, while members appear as blue dots. Each new center is $30''$ from the original center. \textit{Top right panel:} Slopes $d(\mu_{r})/dr$ obtained from a linear fit in the region of $1'$--$5'$ from the cluster center as in Fig.~\ref{radial_PM_fit}. Slopes are computed for the different centers located at different direction angles, as shown in the left panel. $0\degree$ angle is the circle located just to the left of the empty star, and all the other $8$ shift direction angles $\theta$ are measured from it and up to $315 \degree$. \textit{Central right panel:} Same as top right panel, but now for the intercepts $\mu_{r}(r=0)$ obtained in the region between $1'$--$5'$ from the center. \textit{Bottom right panel:} Same as top and central right panels, but now for the intercepts in the horizontal axis $r(\mu_{r}=0)$ obtained in the region between $1'$--$5'$ from the center. The error bar in the right panels illustrates the standard deviation $\sigma$ of the values in the vertical axis.}
    \label{centers}%
\end{figure*}

Since the cluster center used to compute $\mu_{r}$ directly affects the results, the expansion signal requires further testing. Adopting a center that deviates from the true physical center of mass may bias our measurements. We therefore verified whether the expansion signal persists when alternative cluster centers are assumed. As shown in Fig.~\ref{PM_color} and Fig.~\ref{PM_phi}, the considered center (black empty star) may be slightly off the location of very massive Wd1 members (central region devoid of star counts). Thus, we moved the center $8$ times around this prior location, as shown in the left panel of Fig.~\ref{centers}. Each new center appears as a red open circle located at a radial distance of $30''$ from the black empty star.\\

To confirm if the expansion is independent of the value of the center considered for computing the $\mu_{r}$ values, after estimating these new values, we fitted a line in the same region from the center that we considered in Fig.~\ref{radial_PM_fit}, between $1'$--$5'$. The slopes obtained with the $8$ new centers are shown in the top-right panel of Fig.~\ref{centers}, whereas the vertical intercepts $\mu_{r=0}$ of the same linear fit are shown in the middle-right panel. The mean value of the computed slopes is $\left< d\mu_{r}/dr\right> = 0.08 \pm 0.02$ mas yr$^{-1}$ arcmin$^{-1}$. Its error represents the standard deviation of the obtained values. For the computed vertical intercepts, the mean value corresponds to $\left<\mu_{r=0}\right> = -0.19 \pm 0.06$ mas yr$^{-1}$. On the other hand, for the horizontal intercepts (bottom right panel of Fig.~\ref{centers}), their mean value was found as $\left<r(\mu_{r}=0)\right>=2.55 \pm 0.25$ arcmin. Given their dispersion errors, all three values appear very stable and independent of the adopted cluster center. In addition, given that all computed slopes have positive values, the expansion effect is also observed regardless of the center considered for the $\mu_{r}$ calculation.\\

Given this center-shift test, we measure an expansion gradient $d\mu_{r}/dr = 0.06 \pm 0.02$ mas yr$^{-1}$ arcmin$^{-1}$ (bootstrap), with an additional centering systematic of $0.02$ yr$^{-1}$ arcmin$^{-1}$ estimated from a $30''$ center-shift test; the quadrature total uncertainty is $0.03$ yr$^{-1}$ arcmin$^{-1}$.\\

\subsection{Asymmetric expansion}

\begin{figure*}
    \centering  \includegraphics[width=0.85\textwidth]{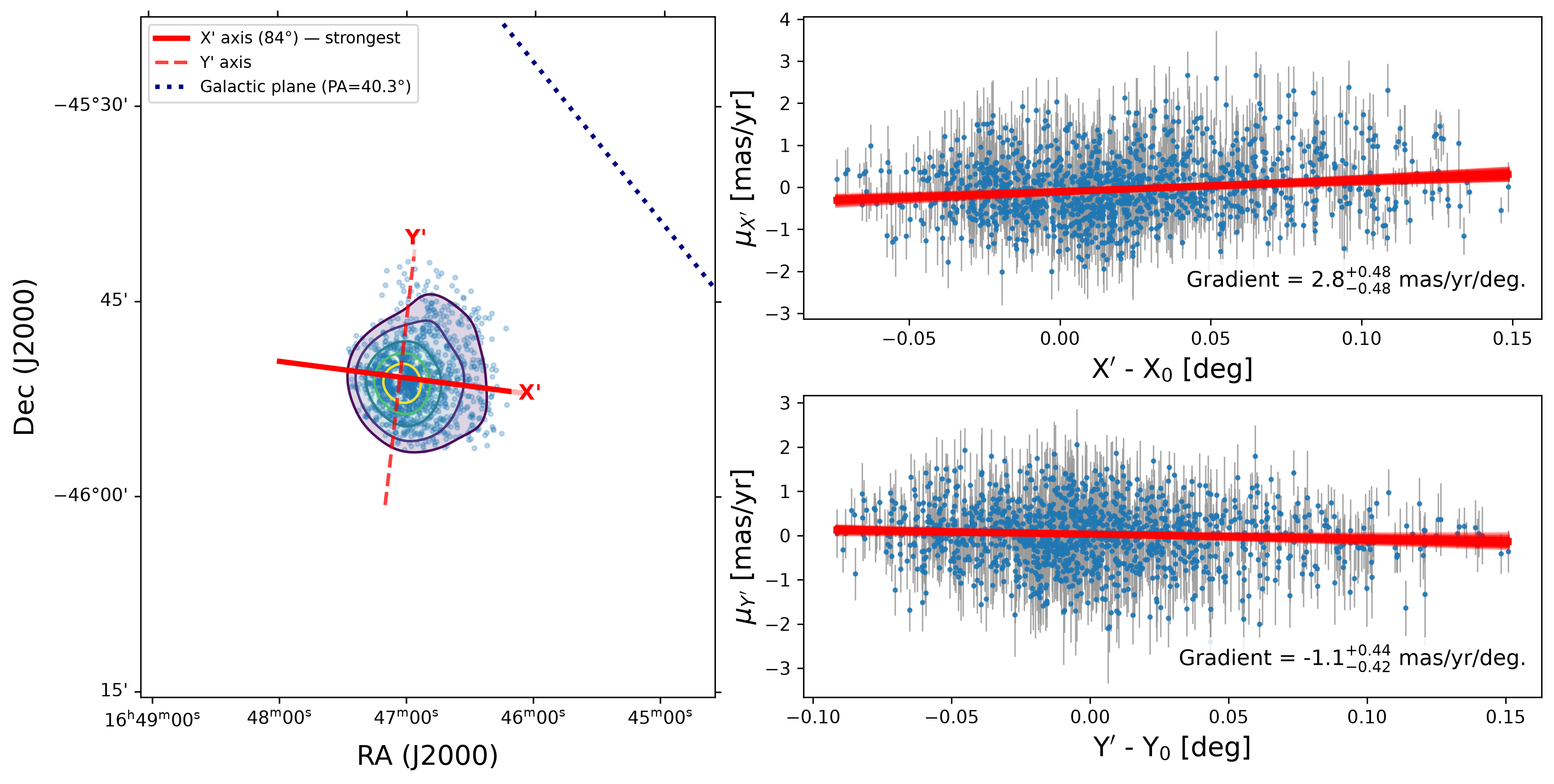}
   \caption{\textit{Left panel:} Spatial distribution of our candidate members (blue dots) with density contours overplotted. We also show the rotated axes $X'$ (dashed red line) and $Y'$ (continuous red line) for comparison. The $Y'$ axis shows the strongest expansion gradient, at a significance of $5.9\sigma$. The dotted dark blue line marks the location of the MW plane. \textit{Top right panel:} PM values (blue dots) and their uncertainties (gray lines) projected along the $X'$ axis, $\mu_{X'}$ against the position of the stars relative to the center of the cluster. \textit{Bottom right panel:} Same as the top panel, but now for the orthogonal axis $Y'$. In both, the red lines show the posteriors of the distribution.}
    \label{XY_pms}%
\end{figure*}

Since several open clusters show asymmetric expansion, we followed \citet{Wright_2018} and \citet{Wright_2024} to check if this effect also occurs in a supermassive star cluster, such as Wd1. To search for the direction in the plane of the sky along which the expansion signature is strongest, we rotated the coordinate system in steps of $1\degree$, defining at each step two orthogonal axes, namely $X'$ and $Y'$. For each rotation angle, we projected the PM vectors onto these axes (and propagated their uncertainties, including correlations between $\mu_{\alpha*}$ and $\mu_{\delta}$). We then performed two independent 1D linear fits, $\mu_{X'}=A_{X'}X'+B_{X'}$ and $\mu_{Y'}=A_{Y'}Y'+B_{Y'}$, modeling the data with a Gaussian likelihood that includes an intrinsic-scatter term to the measurement errors. We sampled the posterior distributions of $A$, $B$, and the intrinsic scatter using a Monte Carlo Markov Chain (MCMC) ensemble sampler \texttt{emcee} \citep{Foreman-Mackey_2013}, and the best rotation angle was selected as the one maximizing $A/\sigma_{A}$ in either axis. We find the most significant expansion gradient at the $X'$ axis, corresponding to PA$=84\degree \pm 8\degree$, shown as a red continuous line on the left panel in Fig.~\ref{XY_pms}. The reported error was estimated using $200$ Monte Carlo realizations, including a bootstrap resampling and PM perturbations based on their uncertainties. To each new modified sample, we searched its PA of strongest expansion. Using this procedure, we found that the median PA of all $200$ realizations was around $84\degree$, and we adopted its $16$th–$84$th percentile interval as its uncertainty. The median significance of the strongest recovered gradient was $4.7\sigma$.\\

Considering the strongest gradient, the projected PMs for each of the $X'$ and $Y'$ axis are shown in Fig.~\ref{XY_pms} against their position $(X, Y)$ relative to the center of the cluster $(X_{0}, Y_{0})$. The upper panel shows the PM values in our rotated $X'$ axis, namely $\mu_{X'}$, as blue dots with their errors and the same is shown for our $Y'$ axis in the bottom panel. The red lines are the posteriors of the distribution. The gradient, or mean of all the slopes, along the $X'$ direction (at PA$=84\degree$) is positive and has approximately $5.9\sigma$ significance. The positive slope and the high significance of the gradient suggest that expansion is stronger in this direction, nearly east to west, and is asymmetrically developed. On the other hand, we found a negative gradient on the $Y'$ axis, which could suggest a contraction. However, this is only at approximately $2\sigma$ significance, so our data do not clearly confirm this effect.\\

\section{Discussion}
\label{discussion}

We must interpret the kinematic patterns observed in Wd1 in the context of the cluster's stellar population properties. Paper I showed that Wd1 exhibits a core–halo spatial structure among intermediate-mass stars, and the studies by \citet{Gennaro_2011} and \citet{Lim_2013} found signs of mass segregation. These constraints limit the range of viable formation scenarios and provide an essential framework for interpreting the expansion and contraction patterns identified here, as we discuss below.\\

\subsection{Expansion and consequences for the formation of Wd1}

Studies of young clusters show that expansion is common among open clusters. In the work by \citet{Kuhn_2019}, the authors analyzed young stars located in $28$ open clusters and associations of the MW with ages between $1$--$5$ Myr. Their analysis showed, through Gaia DR2 PMs \citep{Lindegren_2018}, that $75\%$
of them are expanding. In addition, \citet{Wright_2024} considered $18$ stellar clusters and OB associations and found that nearly all are expanding. The latter analysis used optical astrometry from Gaia \citep{Gaia_2023} and spectroscopy from the Gaia-ESO Survey. They found 3D velocity dispersions to be anisotropic, confirming these clusters were not dynamically relaxed.\\

The expansion observed in Wd1 confirms that the cluster is not dynamically relaxed, a behavior that, while expected from a dynamical point of view, has not previously been reported in a supermassive star cluster. From NIR PMs of its members, this expansion is most clearly detected in the outer regions of the cluster. Although radial velocity measurements would strengthen this result, our data already indicate expansion. This expansion may also reflect a cloud dispersal process driven by feedback from massive stars \citep{Lada_1984}.\\

In addition, another possible origin of the asymmetry observed in the expansion pattern of Wd1 is related to its formation. First, if the cluster formed through a cloud-cloud collision and the stars retain the kinematic information of the collision, then the expansion develops preferentially along the collision axis. \citet{Furukawa_2009} have already claimed this type of process to be responsible for the formation of Wd2 and, more recently, \citet{Sano_2026} found evidences of triggered star formation through a cloud-cloud collision process in Wd1. On the other hand, the observed asymmetric expansion can also be linked to subcluster mergers that developed in a certain direction and formed the current Wd1 cluster \citep{Wright_2019}. Given our current data, we can not discard any of the formation mechanisms of the cluster considered here. Further analysis, including measurements of the velocity of nearby clouds, can shed light on this issue.\\

Since a narrow age spread has been observed among the members of Wd1 (see e.g. \citeauthor{Gennaro_2011}, \citeyear{Gennaro_2011}; \citeauthor{Kudryavtseva_2012}, \citeyear{Kudryavtseva_2012}), we expect the cluster formed within a short timescale, favoring either a monolithic or a rapidly assembled hierarchical scenario (nearly monolithic). At the same time, the core–halo stellar structure observed in our member sample suggests that Wd1 may have assembled from the early merger of several subclusters. Such mergers must have occurred rapidly and involved coeval stellar populations, given the small age spread of the cluster. This scenario is consistent with a nearly monolithic formation process, in which substructures merge efficiently at very early stages to form a centrally concentrated system \citep{Banerjee_2015}. The expansion observed in the outer regions, together with the asymmetric shape of the cluster, indicates that Wd1 is not fully dynamically relaxed. In contrast, the possible contraction of the inner regions may trace the continued settling of material toward the cluster center after rapid early assembly. In this context, any observed mass segregation would not by itself prove a subcluster-merger origin, but would be consistent with accelerated early dynamical merging.\\

The axis in which asymmetric expansion we found to be strongest is not consistent with the elliptical morphology of the cluster reported by \citet{Gennaro_2011} and \citet{Negueruela_2022} from NIR and optical data, respectively. In our work, we found that expansion is stronger at a PA$=84\degree \pm 8\degree$, which differs from the elongation direction of the cluster observed through the diffuse X-ray emission by \citet{Muno_2006} with an inclination of $13 \pm 3\degree$ east from north. \citet{Wei_2025} found that the cluster is elongated in a position angle of PA $\sim 35.2$ east of north, which also differs from our direction of strongest expansion PA, but it is in close agreement with the PA of the bulk motion of the cluster. Considering the mean PM value of our members ($\overline{\mu_{\alpha*}}=-2.36 \pm 0.02$ mas$/$yr and $\overline{\mu_{\delta }}=-3.70 \pm 0.02$ mas$/$yr), the bulk motion is developed at PA$\approx 32.5\degree$ west from south. Additionally, our obtained PA for the strongest expansion is in close agreement with the PA of the anisotropic expansion found by the ongoing work by Prisinzano et al. (in prep) using Gaia DR3 data. We note that all the studies mentioned probe different stellar populations across different mass ranges and cluster components, and use a variety of spatial extents, so discrepancies are expected.\\

\subsection{Asymmetric stellar distribution}

The spatial distribution of the Wd1 members found in Paper I is asymmetric, with a halo of stars towards the north-west of the cluster center. In the opposite direction, there is a lack of stars, producing this uneven stellar distribution. We tested whether this reflects the true distribution of the Wd1 members or whether it is dominated by differential extinction using the map by \citet{Marsh_2017}. As mentioned in paper I, VIRAC2 data appear to be only marginally affected by this highly extinct belt since, by taking into account the map of warm dust by \citep{Marsh_2017} and the star counts detected by VIRAC2 in two different dust emission zones (one with $4\times$ higher emission than the other), we found almost the same number of objects in both. The latter suggests that the observed stellar distribution could be the true one.\\

We also note that the stellar halo is oriented towards the MW plane, as shown by the dark blue dotted line in the left panel of Fig.~\ref{XY_pms}. Tidal interactions between the Galactic plane and the cluster can affect its stellar distribution \citep{Binney_2008}. If the cluster crossed the plane during its lifetime, this would increase the likelihood of such an asymmetric distribution \citep{Martinez-Medina_2017}. Considering that the cluster center lies $0.4\degree$ below the MW plane and using the mean PM value of the cluster members reported in Paper I, we estimate how long ago the cluster may have crossed the plane. Assuming Wd1 moves at a constant speed in a straight line, the cluster would have crossed the MW plane $\approx 2.35$ Myr ago, which is shorter than its age. The asymmetric stellar distribution could therefore be associated with this process, although further analysis is needed to confirm its origin. The fact that the elongation PA found by \citet{Wei_2025} is in agreement with the PA of the bulk motion of our members further supports the hypothesis that the passage of the cluster through the galactic plane may have given rise to its elongated shape.\\

The $2.35$ Myr value obtained above can be tested by a dynamical approach. Considering the Galactic disk and halo potentials \citep{Miyamoto_1975, Navarro_1996}, assuming a galactocentric distance of Wd1 of $4$ kpc, and computing the vertical oscillation parameters, we found that Wd1 should have crossed the MW plane $2.4$ Myr ago, which is in close agreement with our first result. The latter analysis also suggests that Wd1, given its $\sim6$ Myr age, must have formed at a vertical distance of $\approx17$ pc above the Galactic plane.\\

\subsection{Possible contraction of the inner parts of the cluster}

\begin{figure}
    \centering  \includegraphics[width=0.45\textwidth]{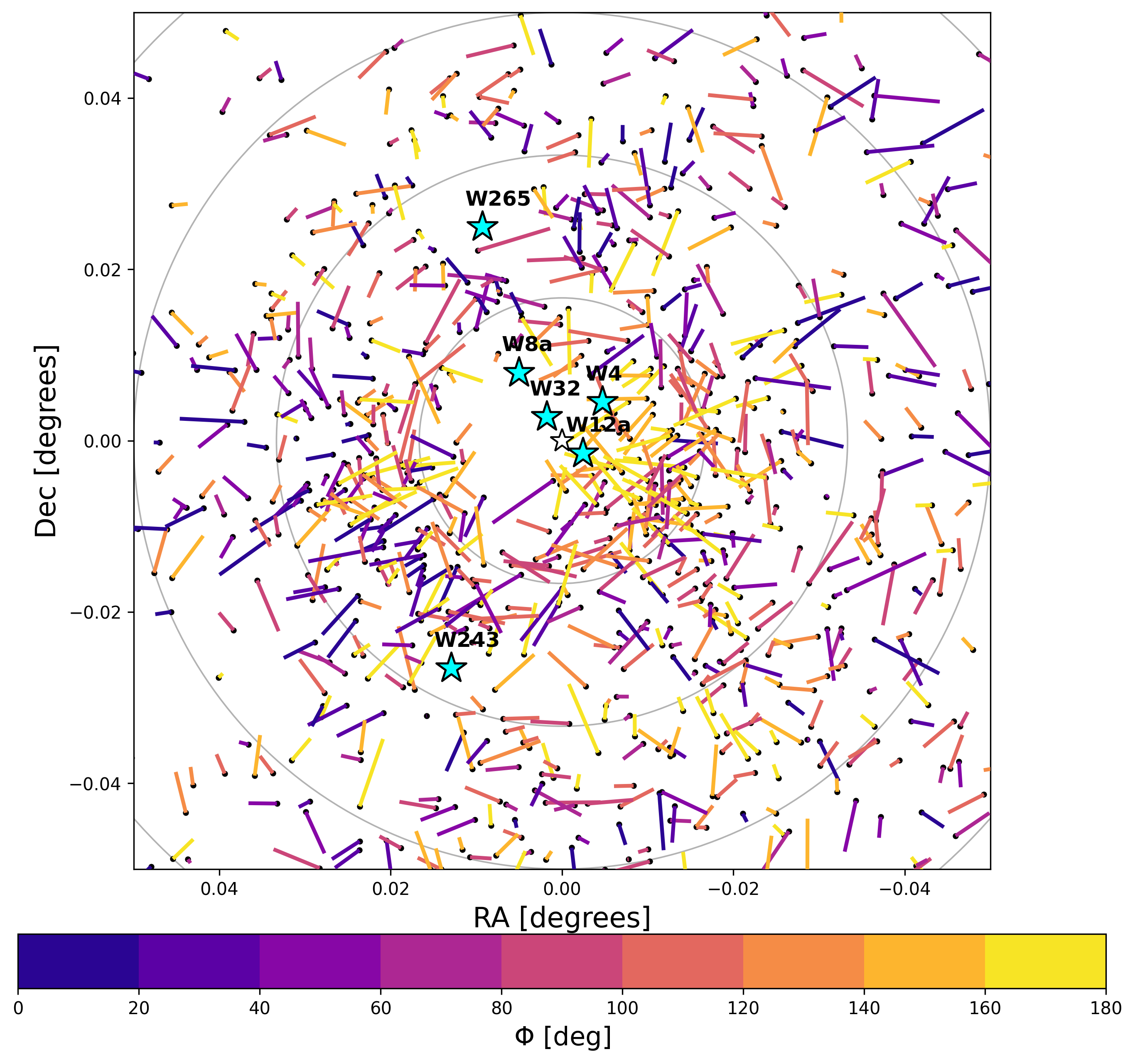}
   \caption{Zoom into the inner $3'$ radius region of Wd1, similar to Fig.~\ref{PM_phi}. The black empty star illustrates the original center of the cluster, while cyan stars show the location of post-Main Sequence sources considered in the work by \citet{Cottaar_2012} for computing the radial velocity dispersion of Wd1.}
    \label{cottaar}%
\end{figure}

The work by \citet{Cottaar_2012} argued that Wd1 is a gravitationally bound system. Using radial velocity measurements of $5$ yellow hypergiants and $1$ luminous blue variable, the authors showed that the radial velocity dispersion was consistent with the cluster being subvirial. Fig.~\ref{cottaar} shows a further zoom into the inner $3'$ radius region, similar to Fig.~\ref{PM_phi}. The sources considered for measuring radial velocity dispersion in \citet{Cottaar_2012} are shown as cyan stars. These sources are predominantly located in the central regions of Wd1, where our NIR PM data suggest a contraction pattern (yellow and orange lines directed toward the center), while the outer regions of the cluster show expansion.\\

\citet{Gennaro_2011} and \citet{ Lim_2013} show that Wd1 is experiencing mass segregation, a dynamical process in which massive stars sink toward the cluster center through gravitational encounters. The latter are expected, given the high stellar density of the core of the cluster \citep{Clark_2005}. Our observation of central contraction, up to $\approx 2.5$ arcmin from the center given the horizontal intercept of Fig.~\ref{radial_PM_fit}, is consistent with this ongoing mass segregation process, the subvirial state found in \citet{Cottaar_2012}, and the mass segregation found by \citet{Lim_2013} in the inner $\approx 1.7$ arcmin from Wd1's center.\\

The simultaneous observation of central contraction and outer expansion in Wd1 reveals its complex dynamical state. The central contraction can be naturally explained by mass segregation, which concentrates massive stars toward the center. The outer expansion has a different origin. In young massive clusters, gas expulsion driven by stellar feedback can render them unbound or loosely bound, leading to expansion \citep{Stahler_2018}. The latter suggests that the cluster may still be responding to early gas expulsion while simultaneously undergoing dynamical mass segregation. Nevertheless, the segregation of massive stars toward the cluster center and their strong winds can lead to cluster expansion. This effect can be further enhanced by supernova explosions, which can drive stronger central mass loss \citep{Banerjee_2017}. Given the presence of a magnetar in the Wd1 region \citep{Muno_2006}, this is also a probable scenario.\\

In a broader evolutionary context, young massive clusters like Wd1 are thought to be the progenitors of globular clusters \citep{Portegies_Zwart_2010, Marks_2012}. The dynamical evolution of such systems is driven by stellar encounters that redistribute kinetic energy between stars of different masses; thus, massive stars lose energy to their lower-mass counterparts, causing them to sink toward the cluster center in a mass segregation process, while the outer regions expand \citep{Spitzer_1987}. In dense, young super star clusters, this can give rise to a runaway regime of stellar collisions and to core collapse. The combination of central concentration and global expansion observed in Wd1 is therefore consistent with an early phase of this dynamical sequence, potentially linking mass segregation to the onset of core collapse. While our analysis does not address the long-term survival of the cluster, understanding this phase is relevant for constraining the early conditions that determine whether young massive clusters remain bound and eventually evolve into systems resembling present-day globular clusters, or instead disperse.\\

\section{Summary and conclusions}
\label{conclusions}

The proper motions of our proposed members of Westerlund~1 are in good agreement with those obtained from Gaia DR3. In addition, we compared our member selection with that of \citet{Wei_2025}, who used HST data to derive photometric and kinematic properties of stars in the central region of Westerlund~1. They identified cluster members primarily through kinematic information combined with color-magnitude diagram positions. This comparison showed that $250$ out of $374$ common sources ($\approx 67\%$) are classified as members in both analyses. The main discrepancies arise for faint sources, likely due to differences in the adopted cluster age and distance in the two studies. Overall, this comparison provides an external validation of our proper motion measurements and member selection.\\

With our proper motions externally validated, we performed a kinematic analysis and found indications of expansion among Westerlund~1 members. Using relative near-infrared proper motions, we found that stars located in the outskirts of the cluster predominantly move away from the cluster center. In particular, within the inner $1'$--$5'$ from the center, a linear fit of the radial component of the relative proper motions, $\mu_{r}$, as a function of distance from the cluster center yields a positive slope of $0.06 \pm 0.02$ mas yr$^{-1}$ arcmin$^{-1}$, consistent with global expansion.\\

To test the robustness of this result, we repeated the analysis using different assumed cluster center positions. The inferred expansion signal remains stable, demonstrating that the observed expansion is independent of the adopted center. In addition, we find that the expansion is asymmetric, with a preferential direction at PA $=84\degree \pm 8\degree$ detected at a significance level of $5.9\sigma$, in contrast to the elongation direction of the cluster found for massive members by \citet{Negueruela_2022} and \citet{Wei_2025} or the diffuse X-ray emission of the cluster found by \citet{Muno_2006}, but in agreement with the strongest velocity dispersion found by Prisinzano et al. (in prep.) using Gaia DR3 data.\\

We also found tentative evidence for contraction in the inner regions of the cluster, but only at a significance level of $\sim 2\sigma$. This contraction was likewise tested against different center positions and found to be stable. The central contraction, extending to $\approx 2.5'$ from the cluster center, is consistent with ongoing mass segregation and the subvirial state reported by \citet{Cottaar_2012}, suggesting that different physical processes dominate at different radii and unveiling a complex dynamical state for Westerlund~1. This can also be related to what has been reported in the literature through simulations, in which an expanding cluster can still retain a virial or subvirial core \citet{Kroupa_2001}.\\

The observed expansion of Westerlund~1 can be associated with the expulsion of residual gas from the parental molecular cloud, potentially enhanced by at least one supernova event. The asymmetric nature of this expansion may reflect the initial conditions of cluster formation, such as a cloud–cloud collision or the early merger of subclusters along a preferred direction. The latter scenario is further supported by the observed mass segregation and inner contraction of the cluster \citep{Gennaro_2011}. Taken together, these results are consistent with a nearly monolithic formation scenario, in which early substructure merges rapidly to form a centrally concentrated system.\\

The asymmetry observed in the stellar distribution of the members could be the result of the passage of Westerlund~1 across the plane of the Milky Way. By considering the mean PM of the cluster members and the distance of the cluster to the Galaxy's plane in the latitude direction, we found that Westerlund~1 should have crossed the plane about $2.35$ Myr ago. This value is lower than the age of the cluster, which is between $5$--$6$ Myr, and could account for its uneven stellar distribution. In this case, its members should have experienced tidal interactions with the Milky Way potential, an origin that still needs to be tested to be confirmed.\\

Further kinematic studies of Westerlund~1 members are required to confirm and extend these results, particularly by probing mass ranges not accessible with the VVVX/VIRAC2 catalogs. In this context, the present work complements Paper I by extending the dynamical analysis of Westerlund~1 and contributes to bridging the gap between different stellar mass regimes.\\

\begin{acknowledgements}
      This work has been supported by FONDECYT regular 1230731, ANID BASAL Center for Astrophysics and Associated Technologies (CATA) FB210003 and by the ANID Millennium Science Initiative, ICN12\_009 and AIM23-0001, awarded to the Millennium Institute of Astrophysics (MAS). COH acknowledges Agencia Nacional de Investigación y Desarrollo (ANID) through FONDECYT postdoctoral grant 3260854. AB acknowledges support under Germany’s Excellence Strategy through the Cluster of Excellence ORIGINS EXC–2094–390783311. RB acknowledges support by the INAF Mini-Grant “Physical properties of Accreting young stellar objects: exploration of their light Curves and Emission (PACE)”. KM acknowledges support from the Fundação para a Ciência e a Tecnologia (FCT) through the CEEC-individual contract 2022.03809.CEECIND, and grant UID/04434/2023. This research was supported by the International Space Science Institute (ISSI) in Bern, through ISSI International Team project \#25-639. This work has made use of data from the European Space Agency (ESA) mission
{\it Gaia} (\url{https://www.cosmos.esa.int/gaia}), processed by the {\it Gaia}
Data Processing and Analysis Consortium (DPAC,
\url{https://www.cosmos.esa.int/web/gaia/dpac/consortium}). Funding for the DPAC
has been provided by national institutions, in particular the institutions
participating in the {\it Gaia} Multilateral Agreement.
\end{acknowledgements}


\bibliography{Wd1_bibliography}

\end{document}